\documentstyle[aps,prl,multicol,epsf]{revtex}

\begin{document}
\def\CC{{\rm\kern.24em \vrule width.04em height1.46ex depth-.07ex
\kern-.30em C}}
\def\P{{\rm I\kern-.25em P}}
\def\RR{{\rm
         \vrule width.04em height1.58ex depth-.0ex
         \kern-.04em R}}

\draft
\title{Universal control of quantum subspaces and subsystems}
\author{{ Paolo Zanardi$^{1,2,3}$ and Seth Lloyd$^1$ }}
\address{
$^1$ Department of Mechanical Engineering,
Massachusetts Institute of Technology, Cambridge Massachusetts 02139 
\\
$^2$  Institute for Scientific Interchange  (ISI) Foundation,
Viale Settimio Severo 65, I-10133 Torino, Italy\\
$^3$ Istituto Nazionale per la Fisica della Materia (INFM)
}
\date{\today}
\maketitle
\begin{abstract}
We describe a broad dynamical-algebraic framework 
for analyzing the quantum control properties of a set of naturally
available interactions. General conditions under which 
universal control is achieved over a set of subspaces/subsystems
are found.  
All known physical examples of universal control
on subspaces/systems  are related to the framework developed here. 
Implications for quantum information processing
are discussed. 
\end{abstract}

\begin{multicols}{2}

The ability to manipulate information in an arbitary fashion
is a key requirement for both classical and quantum information processing (QIP) \cite{QC}.
Once information is suitably encoded one must be able 
to perform, at least approximately, any transformation over the state space
of the physical medium supporting the encoding.  When this goal is realized
one says that {\em universality} is achieved.

In the protoype case of QIP the physical system supporting the encoding
is provided by a set of two-level i.e., qubits, in which both 
external and mutual interactions are supposed to be  controllable 
to a very high degree of accuracy. In this case the state-space
of the systems is given by the tensor product  ${\cal H }\cong (\CC^2)^{\otimes N}$ (N-qubit space).
It is an important, and by-now standard result in QIP that almost any two-qubit gate is universal
\cite{SETH},\cite{UG}.  Moreover the realizability of
all single-qubit i.e., $SU(2)$ gates along with the one of an (arbitary) entangling two-qubit gates
suffices to achieve universality \cite{mike}.

On the other hand in many experimental  situations there are operational constraints  
that force one to consider a smaller set of transformations as the  actually available ones.
For example all naturally available interactions could be commuting with some observable
e.g., total spin, whose value cannot then be changed.
This lack of resources typically results in the impossibility of achieving universality
in the full state space  $\cal H.$ It is then a very natural and practically important question
whether there exists a subspace $\cal C$ of $\cal H$ over which  the restricted set of
naturally available interactions allows universality.
When such an ``encoding'' is  found one obtains the so-called
{\em encoded universality}  \cite{danFT,danNS,enc,Dan,lorenza}.

In this paper we shall analyze the problem of encoded universality from
a  general  control-theoretic perspective. Broad conditions
under which  universal control over set of subspaces/subsystems
can be achieved will be stated within
 powerful algebraic framework.   The main actors of the latter
will be the dynamical groups and algebras associated with the allowed
interactions. A crucial role will be  played by the   symmetry properties
of the realizable transformations. Several applications to physical  systems 
relevant for quantum information processing will be pointed out.

{\em Preliminaries.}--
Let ${\cal I}_A:=\{ H(\lambda)\}_{\lambda\in {\cal M}}\subset \mbox{End}({\cal H})$
denotes the set of `` naturally'' available interactions acting over the
quantum state-space ${\cal H}.$
$\cal M$ is the set of control parameters.
We assume that one is able to enact all the quantum evolutions governed by the time-dependent
Hamiltonians $H(\lambda(t))$ where $\lambda\in{\cal P}_A$ is the set
 of ${\cal M}$-valued functions (paths) of time corresponding to the physically realizable control
 processes.   We stress that we are not assuming that these latter can be arbitray ones i.e., 
that ${\cal P}_A={\cal F}(\RR,{\cal M}).$ 

The {\em pair} $({\cal I}_A, {\cal P}_A)$ describes the physical resources available in the given 
experimental situation; associated with it one has a  set of allowed 
quantum evolutions 
$U(\lambda)=T\exp ( -i\int_\RR H(\lambda(t)) dt ) \quad (\lambda\in {\cal P}_A)
$

We will assume that if $U$ is an allowed evolution, then $U^\dagger$ is allowed as well;
we also assume that the trivial i.e., $U=\openone,$ evolution is an allowed one. 
It follows that set of unitary transformations one can generate by resorting to
interactions in ${\cal I}_A$ 
and control processes in ${\cal P}_A$ has the structure of a {\em subgroup} ${\cal U}_A$ of
${\cal U}({\cal H}).$
If ${\cal U}_A$ is dense in ${\cal U}({\cal H})$ one says the {\em universality} is achieved:
an arbitrary unitary transformtion over ${\em H}$ can be realized to an arbitrary accuracy
by means of the available  resources. 

It is useful now to recall  a well-known result in quantum control theory. 
When (i)
${\cal P}_A={\cal F}(\RR,{\cal M})$ i.e., one can drive the control parameters along arbitary paths 
in ${\cal M}$  and (ii) ${\cal I}_A=\{\sum_i \lambda_i H_i\}$, one  has that
\begin{equation}
{\em U}_A=e^{ {\cal L}_A }
\label{ctrl}
\end{equation}
where by ${\cal L}_A$ we denoted the Lie algebra generated by the set of operators ${\cal I}_A$
i.e., the linear span of all possible multiple commutators of elements of ${\cal I}_A.$ 
This result generally {\em does not} hold
when a restricted set of paths ${\cal P}_A$ is
considered: in this case ${\cal U}_A\subset e^{{\cal L}_A}. $

For example, in Holonomic
Quantum Computation (HQC) \cite{HQC} ${\cal I}_A$ comprises a set of iso-degenerate Hamiltonians 
and  ${\cal P}_A$ is given by adiabatic {\em loops} around a $\lambda_0\in{\cal M}.$
From the adiabatic theorem it follows that, if one start from an initial state lying
in a eigenspace of $H(\lambda_0),$ any evolutions obtained by driving the control parameter adiabatically 
along a loop in $\cal M$ will result in a final state in the same eigenspace.
This means that the state space is dynamically decoupled in orthogonal sectors corresponding
to the eigenprojectors of $H(\lambda_0).$ This decoupling is clearly an obstruction to universality.

{\em Encoded Universality.}--
Suppose that there exist a set of invariant subspaces 
${\cal C}_i\subset {\cal H}\,(i=1.\ldots,M)$ of ${\cal U}_A,$ such that
\begin{equation}
{\cal U}_A|_{{\cal C}_i}= {\cal U}({\cal C}_i),\quad (i=1.\ldots,M).
\end{equation}
In this case, we say that  ${\cal U}_A$ is  ${\cal C}_i$-universal. The ${\cal C}_i$'s 
will be referred to as codes. When  ${\cal U}_A$ is ${\cal H}$-universal we will simply say
that it is universal. Notice that  in order to attain $\cal C$-universality
the group ${\cal U}_A$ has to be an {\em infinite}  one.  Finite groups cannot
be dense on the set of unitary transformations on ${\cal C}_i$.  

{\em Example 0.}
The most favorable case of holonomic quantum computation occurs when  
there is an irreducible connection \cite{HQC}.  In this case, one has
${\cal U}_A=  \oplus_r {\cal U}({\cal H}_r)$ where ${\cal H}_r$ is the $r-th$ 
eigenspace of $H(\lambda_0)$ with dimension $n_r.$
Since for non-trivial $H(\lambda_0)$ one has
$\sum_r n_r^2 < (\sum_r n_r)^2$ --  ${\em U}_A$   is strictly contained in ${\cal U}({\cal H}).$
Here ${\cal U}_A$ allows only for ${\cal H}_r$-universality.  

{\em Example 1}
Let ${\cal H}=\CC^2\otimes \CC^2$ a two-qubit space and
${\cal I}_A=\{ \sigma^x\otimes\sigma^x+ \sigma^y\otimes\sigma^y, 
\sigma^x\otimes\sigma^y- \sigma^y\otimes\sigma^x, \sigma^z\otimes \openone -\openone\otimes\sigma^z\}
.$ Under the assumptions for the validity of Eq (\ref{ctrl})
it is easy to see that this set  
is ${\cal H}_1$-universal, where ${\cal H}_1$ is the linear span of $|01\rangle$ and $|10\rangle$
\cite{Dan}.
This is easily seen by noticing that $({\cal L}_A)\cong su(2);$  consequently
${\cal H}$ splits
according the $su(2)$ irreducible representation (irrep)
 in a triplet (${\cal H}_1$) and two singlets (${\cal H}_0$).
The decomposition of the entire two-qubit space is obtained by 
considering
${\cal I}_A^\prime=\{ \sigma^x\otimes\sigma^x- \sigma^y\otimes\sigma^y, 
\sigma^x\otimes\sigma^y+\sigma^y\otimes\sigma^x, \sigma^z\otimes \openone +\openone\otimes\sigma^z\}
$.  In this case, the role of ${\cal H}_0$ and ${\cal H}_1$ are interchanged.

It is important  to realize that in the general case the codes
 do {\em not} have to be ${\cal I}_A$-invariant subspaces; in other words, one
can temporarily leave the coding subspace during the time-evolution  and return
to it just at the end. An instance of this situation is provided by the obvious fact
that if $({\cal I}_A, {\cal P}_A)$ is ${\cal C}$-universal then, for any subspace
${\cal C}^\prime\subset{\cal C},$ there exists a subset ${\cal P}_A^\prime\subset{\cal P}_A$
such that $({\cal I}_A, {\cal P}_A^\prime)$ is  ${\cal C}^\prime$-universal.
The elements of ${\cal U}_A$ will generally temporarily draw states out of ${\cal C}^\prime;$
the states in $({\cal C}^\prime)^\perp$ play the the role of {\em auxiliary} intermediate states
that do not have to appear at the beginning  and at the end of the control process.
The QIP literature provides a multitude of illustrations of this state of affairs,
i.e., the use of {\em ancill\ae}. 
Another possibility consists in generating from
the interactions in ${\cal I}_A$ (which do not leave ${\cal C}$ invariant)
a set ${\cal I}_A^{eff}$ of effective interactions (which do leave ${\cal C}$ invariant).

It is interesting to notice that this is the case even in the so-called
topological quantum computation \cite{TOP},\cite{ZL}.
There the code is provided by the ground-state of a many-body Hamiltonian
whose degeneracy arises and it is protected by a broken topological symmetry.
Manipulations of the codewords are then realized by creating anyon-like
excitations, braiding them around in some non-trivial i.e., global, fashion
and returning into the ground-state.

Now the main question is: {\em given the available set ${\cal U}_A$ of operations,
can some encoded universality be achieved?}

To see whether a suitable encoding exists, i.e., a subspace $\cal C$ for wich ${\cal U}_A$
is $\cal C$-universal, it is useful to resort to the tools of group representation
theory \cite{CORN}
Let us consider  the decomposition of ${\cal H}$ according the ${\cal U}_A$-irreps 
\begin{equation}
{\cal H}=\oplus_J \CC^{n_J}\otimes {\cal H}_J
\label{split}
\end{equation}
The $\CC^{n_J}$ factors  in the Eq. above simply take into account that the $J$-th irrep
${\cal H}_J,$
 with dimension 
$d_J,$ appears with multiplicity $n_J.$ The appearance of these factors amounts to the existence
of {\em symmetries} for the set of allowed transformations ${\cal U}_A$.
We  observe in passing that symmetries for ${\cal U}_A$ are not necessarily symmetries for ${\cal I}_A,$
whereas the converse holds true. 

Let us now then suppose that ${\cal I}_A$ admits a non-trivial group of symmetries $\cal G.$
A paradigmatic instance  is given when one is dealing with a quantum system  consisting of $N$ copies of an
elementary one e.g., one qubit, and cannot discriminate the different subsystems.
Permutations of  these latter   are therefore symmetries of  the allowed interactions
($\cal G$ is given  by the symmetric group ${\cal S}_N).$
This kind of situation is often encountered in  Decoherence Free Subspaces (DFS) \cite{EAC}
and noiseless subsystem theory \cite{KLV},\cite{stab},\cite{danNS}.
where ${\cal I}_A$ is the set of system operators coupled with the environment. The  algebra
generated by ${\cal I}_A$ is  the basic algebraic object
underlying all the quantum noise avoidance/correction/suppression schemes developed to  date
\cite{stab}\cite{ZL}. 

In Eq. (\ref{split})  now the $\CC^{n_J}$ factors represent  
the $\cal G$-irreps and $d_J$ their multiplicities.
In this case universality is obviously prevented because ${\cal U}_A\subset \CC{\cal G}^\prime\cong
\oplus_J \openone_{n_J}\otimes M_{d_J}(\CC):$ different $J$ sectors are never coupled
by the allowed operations in ${\cal U}_A.$
In order to better illustrate these notions let us go back to
example 1; here one can choose as symmetry group 
${\cal G}=\{\openone, \sigma^{z\otimes\,2}\}\cong {\bf{Z}}_2.$
Its commutant is then given by 
$(\CC{\cal G})^\prime=\mbox{span}\{\openone,
\sigma^z\otimes\openone,\openone\otimes\sigma^z, \sigma^{z\otimes\,2},
\sigma^\alpha\otimes\sigma^\beta\,(\alpha,\beta=x,y)\}.$
This algebra contains both the $su(2)$'s mentioned above
and it allows one to operate {\em simultaneously} over ${\cal H}_0$ and
${\cal H}_1.$

The group ${\cal U}_A$ acts irreducibly over the subspaces ${\cal C}_J=|\phi\rangle\otimes{\cal H}_J.$
It is elementary, yet important to keep in mind that irreducibility on itself does {\em not}
imply that all the unitaries over $\cal C$ are realized as group elements (see Prop. below).
The most general of such  transformations, as written above, is given by a suitable
{\em linear combination} of elements from  ${\cal U}_A.$ Technically this is expressed by saying that
group of unitaries over $\cal C$ is given by the restriction to $\cal C$ of the unitary part
of the {\em group algebra} of ${\cal U}_A$ \cite{cg} 
i.e., ${\cal U}({\cal C})= U\CC{\cal U}_A|_{\cal C}.$
When the group ${\cal U}_A$ is a Lie one, one can easily prove the following.

{\em Proposition 1}--
If dim ${\cal U}_A|_{{\cal H}_J}=d_J^2-1$ then
${\cal U}_A$ is
${\cal C}_J$-universal where ${\cal C}_J$ is any $d_J$-dimensional
subspace of the form $|\phi\rangle\otimes {\cal H}_J, (|\phi\rangle\in\CC^{n_J}). 
$

{\em Proof.}
From Eq (\ref{split}) it is clear that any of the ${\cal C}_J$ is an irrep space of ${\cal U}_A$
and it is therefore ${\cal U}_A$-invariant.
Moreover under the current assumptions the Lie group ${\cal U}_A$ has dimension $d_J^2-1,$
this means that it coincides with the whole set of (special) unitary transformations over ${\cal C}_J.$
$\hfill\Box$

{ This proposition  provides in principle a protocol for determining whether
a set of  Hamiltonians ${\cal I}_A$ allows for encoded universality:
}
(0) Determine the group ${\cal U}_A$ of allowed unitaries
(1) Decompose the total state-space $\cal H$ according the ${\cal U}_A$ irreducible sectors
(2) compute  for all the $J$'s the numbers $d_J^2 -\mbox{dim}\,{\cal U}_A|_{{\cal H}_J}\ge 0,$
 those equal zero give rise to a $n_J$-parameters family of codes over which ${\cal I}_A$
is universal.
Of course both steps (0) and (1) are in  general  not trivial and represent on their
own a challenge. The situation gets somewhat simplified when the conditions of Eq. (\ref{ctrl}) hold.
In this case everything can be formulated in terms of the Lie algebra ${\cal L}_A.$
In several istances of interest one has that ${\cal L}_A$ is the image of a known Lie algebra $\cal L$
e.g., $su(L)$ though a {\em faithful} i.e.,  zero kernel, irreducible representation $\rho_A.$
In this case $\mbox{dim}\,{\cal L}_A|_{{\cal H}_J}=\mbox{dim}\, {\cal L},$
so it is sufficient to check the  $d_J^2$'s  against a single number e.g., dim $u(2)=4$

{\em Example 2.}
Let us consider $L$ bosonic modes, $[b_i, b_j^\dagger]=\delta_{ij},\,(i,j=1,\ldots,L).$
The set of controllable interactions is given by 
${\cal I}_A=\{ b_j^\dagger b_i\,/\,i,j=1,\ldots,L\}.$
It is a standard matter to see that the bilinears $b_j^\dagger b_i$  span a  algebra ${\cal L}_A$
isomorphic to $u(L)$. 
The Fock space ${\cal H}_F=h_\infty^{\otimes\,L}$ ($h_\infty$ is the state-spae of a single
quantum oscillator) splits in $su(L)$-invariant subspaces ${\cal H}_N$ with dimensions 
$d_{N,L}:=\pmatrix{ & N+L-1\cr &L-1}$ corresponding to the eigenvalues $N$ of the total number operator 
$\sum_{j=1}  b_j^\dagger b_j.$
Typically $d_{N,L}^2> L^2=\mbox{dim}\, u(L)$ and therefore ${\cal L}_A$ is {\em not}
${\cal H}_N$-universal. 
When  $N=1,$ with $L$-arbitrary, one obtains the fundamental irrep for which
$d_{1,L}=L.$ 

{\em Group algebra universality.}--
We illustrate now another general route to encoded universality; particular instances of this
scheme have already found explict important applications in spin-based QIP \cite{enc},\cite{danFT},
\cite{danNS}
and fault-tolerant computation over DFSs \cite{danNS}. 

{\em Proposition 2}--
Suppose  that  the allowed interactions are completely controllable and 
happen to belong to the group algebra
of a non-abelian group $\cal K$ i.e., ${\cal I}_A\subset \CC{\cal K}.$
Then the group ${\cal U}_A$ is {\em generically} $\cal C$-universal for all ${\cal C}=|\phi\rangle\otimes
{\cal H}_J,$ where ${\cal H}_J$ is a $\cal K$-irrep space and   $|\phi\rangle\in\CC^{n_J}$
($n_J$ is the multiplicity of the $J$-th irrep)

{\em Proof.}
Under the current assumptions one has ${\cal U}_A=\exp{{\cal L}_A},$
but for {\em generic} ${\cal I}_A\subset \CC{\cal K}$ one has \cite{SETH}
the Lie algebra generated by the allowed interactions is the {\em whole}
algebra of anti-hermitean elements of the group-algebra $\CC{\cal K}$ 
i.e.,  $u(\CC{\cal K}).$
Thus   ${\cal U}_A|_{\cal C}= \exp{ u(\CC{\cal K})}|_{\cal C}= {U}\CC{\cal K}|_{\cal C}.$ 
But it is a basic fact of group representation theory that the unitary part of the group-algebra
restricted to an irrep-space  amounts  the {\em whole} unitary group over that space.
Formally  ${U}\CC{\cal K}|_{\cal C}={\cal U}({\cal C});$ this relation along with the previous one 
completes the proof. $\hfill\Box$

{\em Example 3}
Let ${\cal H}\cong \CC^2$,  the ${\cal K}:=SU(2)$ fundamental representation space
(one irrep with multiplicity one). A generic Hamiltonian in $\CC SU(2)$
has the form $H=\sum_{\alpha=x,yz} \lambda_\alpha \sigma^\alpha.$
This latter is universal over ${\cal H}.$

At this point it is worthwhile to emphasize that even if both Prop. 1 and 2
have been formulated in terms of subspaces $\cal C$'s simply by tracing out the
$|\phi\rangle$ vectors one gets conditions under which universal control
is achieved over the the factors ${\cal H}_J$ in Eq. (\ref{split}).
The ${\cal H}_J$ factors  correspond to ``virtual'' subsystems in 
which one can decompose the systems according the given available
operational resources \cite{virt}. This kind of quantum subsytem
generalizes the noiseless subsystems \cite{KLV} that form the basis
of general error correction/avoidance strategies \cite{stab},\cite{ZL}. 
It is also interesting to note that Prop. 2 provides us with an example of a group i.e., ${U}\CC{\cal K}$
for which Prop. 1 {\em always} holds true (notice that $\forall J,$ dim$\,{U}\CC{\cal K}=|{\cal K}|>d_J^2$).

As mentioned above, an instance of Prop. 2 is the well-known case of
$N$ spin $1/2$ systems 
coupled by exchange interactions \cite{enc}.  In this case the naturally allowed
Hamiltonian are actually members of the symmetric group  ${\cal S}_N$
(and so are {\em a fortiori} elements of its group algebra).  As a result,
universality can be generically
achieved in any irreducible subspace of the permutation group. For example, for $N=3$
one has one totally symmetric irrep (corresponding to the maximal spin $J=3/2$) 
and a two-dimensional ${\cal S}_3$ irrep (corresponding to two $J=1/2$ $SU(2)$-irreps).
So one has a two-parameter family of encoded qubits over which the exchange Hamiltonians
are universal.

{\em Example  4}
Let us consider as $\cal K$ the simplest non-abelian group: 
the dihedral group $D_3$ \cite{CORN}
i.e.,  the group of spatial rigid symmetries of a triangle
(notice that $D_3\cong {\cal S}_3$). $D_3$ has order six and 
is generated by a $2\pi/3$-rotation 
$R$ and a reflection $P$ satisfying the relations $R^3=P^2=R P R P=\openone.$
A three-dimensional representation is provided by ${\hat R}(z_1,z_2,z_3)=(z_3,z_1,z_2),\,
{\hat P}(z_1,z_2,z_3)=(z_2,z_1,z_3).$ This is a reducible representation: $\CC^3$ splits
in a two-dimensional irrep
${\cal C}\cong\mbox{span}\{\sum_{j=1}^3e^{ {2i\pi}/{3} k\,j} |j\rangle, \,(k=1,2)\}
$
and a one-dimensional irrep  $|s\rangle=1/\sqrt{3}\sum_{j=1}^3|j\rangle$.
The two-dimensional irrep can encode for a qubit.
Now it is easy to check that $P|_{\cal C}=\sigma^x,$ moreover $R-R^{-1}|_{\cal C}$ is proportional 
to $\sigma^z$.
The controllability of
generic hermitean element of $\CC D_3$ e.g., $H(\lambda_1, \lambda_2)=\lambda_1 P + \lambda_2 R
+ \bar \lambda_2 R^{-1}$ then suffices for universal control over ${\cal C}.$

{\em Tensor product structure.}--
Above, it was shown generically how universal quantum control can be obtained
over subspaces/subsystems.  To relate these results to quantum
computation\cite{QC}, we investigate the subcase of quantum control
in which the control space possesses a tensor product structure. 
We then consider a state space ${\cal H }_N={\cal H}^{\otimes\,N}$ 
associated to $N$ copies of a basic one. 
We assume that ${\cal U}_A\subset {\cal U}({\cal H}_N) \supset {\cal U}({\cal H})^{\otimes\,N}$ is 
{\em locally} universal, in the sense that it
contains a sub-group ${\cal U}_{A,loc}^{\otimes\,n}$
such that ${\cal U}_{A,loc}\subset {\cal U}({\cal H})^{\otimes\,M}$ 
is $\cal C$-universal for some ${\cal C}\subset {\cal H}^{\otimes\,M}\,(n:=N/M\in{\bf{N}}) .$
In other words we assume  that there exists a {\em local} encoding, involving a cluster
of $M$ basic subsystems, for which universality is achieved. {\em Example 1} above provides
an instance of this situation in which  two physical qubits are used to encode
a single logical one over which the allowed operations are universal.
Now what one wants is to be universal over the global code ${\cal C}_N= {\cal C}^{\otimes\,N}$.
By the  results in universality contained in Ref \cite{mike}
the following formal result follows.

{\em Proposition 3}--
Let ${\cal U}_A$ be locally universal 
and let there exist $X\in {\cal U}_A$ such that
for any pair $i,j=1,\ldots,M$:
i) $X$ acts as the identity in all the clusters but the $i$-th and the $j$-th;
ii) ${\cal C}^{(i)}\otimes {\cal C}^{(j)}$ 
is an $X$-invariant subspace and  $X$ is an entangling operator over it.
Then  ${\cal U}_A$  is ${\cal C}_N$-universal.

The DFS theory \cite{EAC} provides once again a clear example
of this result. Let ${\cal H}\cong \CC^d$ and suppose that one is able just to turn on and off
exchange Hamiltonians between the different factors in ${\cal H}^{\otimes\,M}.$ 
In this case the available interactions lie in $\CC {\cal S}_M$. 
The commutant of the latter
is given by the $M$-fold tensor representation of $SU(d).$
For $N=2\,d$ the state-space contains a two-dimensional 
$SU(2)$-{\em singlet}  sector $\cal C$, i.e., states in $\cal C$ that are
invariant under all the $SU(d)$ transformations. 
This logical qubit --which requires a cluster of $2\,d$ physical ones --
supports  a $S_M$-irrep \cite{CORN}. Now we consider $n=N/M$ clusters
coupled together by Hamiltonians in $\CC S_N$
(which supports a $S_N$-irrep).
The crucial point is now that the $SU(d)$-singlet sector of ${\cal H}^{\otimes\,N}$
strictly {\em includes}  ${\cal C}^{\otimes\,n}.$
Since  exchange Hamiltonians allow
generically for universality on the former (Prop. 2), 
one gets  ${\cal C}^{\otimes\,n}$-universality as well.
This, in the qubit case $d=2,$  has been constructively shown in \cite{danFT}. 

Even the tensorized form of Example 1 falls
in our scheme. Here, the code is the (tensor power of) the trivial
irrep of group generated by $ie^{i\pi/2\,\sigma^z\otimes\sigma^z}.$ The commutant of this
group --besides all the transformations needed for one-qubit gates --- 
contains elements of the form $\sigma^z_j \,\sigma^z_{j+1},$
which are  used to enact an  entangling  two-qubit gate \cite{Dan}.

{\em Conclusions.}--
In this paper we have formulated  the problem of  universal quantum control and  
quantum information processing on subspaces/subsystems
within  a general algebraic-dynamical
framework. All physical examples known so far
fit in this framework. Constructions have been given providing general conditions under
which encoded-universality can be established. This has been  
done by exploiting the algebraic formalism introduced 
to describe in a unified fashion all known error correction/avoidance
schemes \cite{KLV},\cite{stab},\cite{ZL}.
This unification is on the one hand pretty remarkable in view of the
apparent sharp diversity of the initial physical problems; on the other hand,
the existence of fundamental connection between diverse error compensation
schemes is not totally surprising
once one realizes the {\em duality} between the task of ``not allowing many bad things to happen''
in error correction and `` making as many as good things happen as possible'' in quantum control.

P.Z. gratefully acknowledges financial support by 
Cambridge-MIT Institute Limited and by the European Union project  TOPQIP
(Contract IST-2001-39215)


\end{multicols}

\begin{references}
\bibitem{cg} The group-algebra $\CC{\cal K}$
of a group ${\cal K}$ is the vector space generated by complex
linear combination of elements of ${\cal K}.$
By  extension of the group-multiplication $\CC{\cal K}$
gets the structure of associative algebra.
\end{references}
\end{document}